\documentclass[preprints,article,accept,moreauthors,pdftex]{Definitions/mdpi} 
\firstpage{1} 
\pubvolume{xx}
\issuenum{1}
\articlenumber{5}
\pubyear{2020}
\copyrightyear{2020}
\history{}
\Title{Smart Black Box 2.0: Efficient High-bandwidth Driving Data Collection based on Video Anomalies}

\usepackage[utf8]{inputenc}
\usepackage{amsmath}
\usepackage{bm}
\usepackage{color,soul}
\usepackage{booktabs}
\usepackage{multirow}
\usepackage{verbatim}
\usepackage{graphicx}
\usepackage{subfig}
\usepackage[ruled,vlined]{algorithm2e}
\usepackage{amssymb}
\usepackage{url}
\usepackage{makecell}
\newcommand{\xhdr}[1]{\vspace{5pt} \noindent {\textbf{#1} }}

\newcommand{\ie}{\textit{i.e.}, }
\newcommand{\eg}{\textit{e.g.}, }
\newcommand{\etc}{\textit{etc}, }
\newcommand{\brian}[1]{}
\renewcommand{\brian}[1]{{\color{green} BY: {#1}}}
\newcommand{\rfedit}[1]{}
\renewcommand{\rfedit}[1]{{\color{blue}{#1}}}
\newcommand{\red}[1]{}
\renewcommand{\red}[1]{{\color{red}{#1}}}

\Author{Ryan Feng*, Yu Yao and Ella Atkins}

\AuthorNames{Ryan Feng, Yu Yao and Ella Atkins}

\address{%
University of Michigan, Robotics Ann Arbor, Michigan, USA, 48109\\
brianyao@umich.edu (Y.Y.), ematkins@umich.edu (E.A.)}

\corres{Correspondence: rzfeng@umich.edu (R.F.)}

\abstract{Autonomous vehicles require fleet-wide data collection for continuous algorithm development and validation. The Smart Black Box (SBB) intelligent event data recorder has been proposed as a system for prioritized high-bandwidth data capture. This paper extends the SBB by applying anomaly detection and action detection methods for generalized event-of-interest (EOI) detection. An updated SBB pipeline is proposed for the real-time capture of driving video data. A video dataset is constructed to evaluate the SBB on real-world data for the first time. SBB performance is assessed by comparing the compression of normal and anomalous data and by comparing our prioritized data recording with a FIFO strategy. Results show that SBB data compression can increase the anomalous-to-normal memory ratio by $\sim25\%$, while the prioritized recording strategy increases the anomalous-to-normal count ratio when compared to a FIFO strategy. We compare the real-world dataset SBB results to a baseline SBB given ground-truth anomaly labels and conclude that improved general EOI detection methods will greatly improve SBB performance.}


\begin{document}

\section{Introduction}
Traditional automotive data collection has focused on low-bandwidth vehicle data such as speed and brake status. However, as autonomous vehicles increasingly seem to be the future of transportation~\cite{ENOCH202052}, these data collection methods are becoming obsolete. Modern autonomous vehicles require large-scale collection of high-bandwidth data (\eg video, point clouds) for algorithm development and validation and verification. Deep-learning networks for common autonomous vehicle perception tasks such as object detection~\cite{liu2016ssd,he2018mask,tian2019fcos}, object tracking~\cite{wojke2017simple,choi2015nearonline,Xiang2015LearningTT} and trajectory prediction~\cite{alahi2016social,yao2019egocentric,yao2020bitrap,salzmann2020trajectron++} need significant quantities of high-bandwidth real-world data for effective training and testing.

Finite on-board storage capacity presents a challenge for high-bandwidth data capture that low-bandwidth data logging systems do not encounter. Recently, event data recorders specialized for such high-bandwidth data capture have been explored. The Smart Black Box (SBB)~\cite{yao2018smart,yao2020sbb} is one such system. The SBB uses pre-defined rules for event-of-interest (EOI) detection and computes data value according to the detected EOI. Data value is then used to determine data compression factors and as the basis for prioritized data recording.

This paper expands the Smart Black Box to record high-priority video data and applies the SBB to a real-world driving dataset. Rather than using pre-defined rules for EOI detection as in~\cite{yao2018smart,yao2020sbb}, machine learning-based methods for generalized EOI detection are applied to derive data value. Raw data is grouped into buffers, compressed, and stored in a priority queue in order to discard low-value data as the storage capacity is filled. The SBB is assessed on real-world driving video. We focus on video data due to the ubiquity of cameras as a sensor in automotive applications. 

This paper offers two primary contributions. Firstly, we apply video anomaly detection (VAD)~\cite{yao2019unsupervised,yao2020when} and online action detection (OAD)~\cite{xu2019temporal,yao2020when} as methods for generalized EOI detection on real-world driving video. To estimate data value from combined VAD and OAD outputs, we introduce a hybrid value method based on a weighted sum. Secondly, we present an updated SBB pipeline incorporating VAD and OAD and designed for the real-time recording of dash camera video data. We find that while the SBB improves the collection and retention of high-value data, improved EOI detection methods are needed to realize the full potential of the SBB.

The paper is structured as follows. First, related work is explored in Section~\ref{sec:related_work}. Then, an overview of the original SBB~\cite{yao2020sbb} is presented and the changes we made for application to real-world data are discussed in Section~\ref{sec:preliminary}. Section~\ref{sec:materials_methods} describes our adjusted data classification system, the updated SBB pipeline, and the new value estimation method. Section~\ref{sec:experiments} presents experimental results on a combined real-world dataset and analyzes the performance of our updated SBB. Section~\ref{sec:conclusions} concludes the paper and discusses future work.
\section{Related Work} \label{sec:related_work}
\subsection{Event Data Recorders}
Automotive event data recorders use low-level triggers, such as vehicle impact or engine faults, to log vehicle data leading up to and during anomalous events~\cite{daSilva2014,Gabler2004CrashSA}. However, these systems focus on low-bandwidth data and do not sufficiently address the storage problems posed by high-bandwidth sensors. In the case that the on-board memory is filled, one of two strategies is used. The more common strategy writes data until memory is full, then stops recording data, meaning that the newest data is dismissed. The second strategy uses a circular buffer equivalent to a first-in-first-out (FIFO) queue. In this model, the newest data overwrites the oldest data. Neither of these strategies considers the value of the data being discarded.

High-bandwidth data recorders address this by using prioritized data recording. In the case of~\cite{yao2018smart,yao2020sbb}, valuable data is identified using pre-defined rules for EOI detection. This paper seeks to build on the prioritized recording strategy in~\cite{yao2018smart,yao2020sbb} by applying methods for general EOI detection.

\subsection{Traffic Video Anomaly Detection and Classification}
Several methods exist for identifying EOIs in autonomous vehicles. Pre-defined rules may be applied based on vehicle odometry~\cite{takeda2012,zhao2017,DINGUS20061127} to identify certain EOIs. Other approaches use physiological signals from the driver~\cite{li2016}.
In recent years, deep learning computer vision techniques have been applied to anomaly detection in first-person driving videos~\cite{Chan2016AnticipatingAI,herzig2019spatiotemporal,yao2019unsupervised,yao2020when}. Other works further attempt to classify the type of anomaly occurring in the video, either offline after the video is fully observed~\cite{wang2016temporal,tran2018closer,feichtenhofer2019slowfast} or in real time~\cite{xu2019temporal}. 
These methods present a way for generalized EOI detection based only on dash camera video. As such, we use methods from \cite{yao2019unsupervised} and \cite{xu2019temporal} in the SBB to assign data value in accordance with our focus on video data. Table~\ref{tab:vad_summary} below presents a summary of reviewed anomaly detection methods.

\begin{table}[htbp]
    \caption{{Video-based anomaly detection and action recognition methods.}}
    \centering
    \renewcommand{\arraystretch}{2.0}
    \begin{tabular}{p{0.13\linewidth}|p{0.18\linewidth}|p{0.60\linewidth}}
        \toprule
        \textbf{Method} & \textbf{Aim} & \textbf{Description}\\
        \midrule
        TAD~\cite{yao2019unsupervised} & Anomaly Detection (unsupervised) & Predicts future bounding boxes using RNN encoder-decoders, then takes the standard deviation of predictions as the anomaly score.\\
        DSA-RNN~\cite{Chan2016AnticipatingAI} & Anomaly Detection (supervised) & Uses a Dynamic-Spatial-Attention (DSA) RNN which learns to distribute soft attention to objects and model temporal dependencies of detected cues.\\
        STAG~\cite{herzig2019spatiotemporal} & Anomaly Detection (supervised) & Uses a Spatio-Temporal Action Graph (STAG) network to model the spatial and temporal relations between objects.\\
        \midrule
        TSN~\cite{wang2016temporal} & Action Recognition (offline) & Sparsely samples video snippets and predicts action using RGB and optical flow data.\\
        R(2+1)D~\cite{tran2018closer} & Action Recognition (offline) & Uses a 3D convolutional neural network with separate 2D and 1D convolutional blocks.\\
        SlowFast~\cite{feichtenhofer2019slowfast} & Action Recognition (offline) & Extracts frames from a low frame rate stream to capture spatial information and a high frame rate stream to capture motion.\\
        TRN~\cite{xu2019temporal} & Action Recognition (online) & Simultaneously detects the current action and predicts the action of the following frame.\\
        \bottomrule
    \end{tabular}
    \label{tab:vad_summary}
\end{table}

\subsection{Real-World Driving Datasets}
The increasing popularity of deep-learning methods for self-driving perception tasks has created a demand for high-quality high-bandwidth datasets.
Naturalistic Field Operation Test projects such as \cite{lewis2005nfot,bezzina2015spmd} have been used in the past to gather large amounts of driving data. One such dataset~\cite{lewis2005nfot} uses 100 cars to log nearly 43,000 hours of video and vehicle performance data over a distance of 2,000,000 miles. The more recent Safety Pilot Model Deployment dataset~\cite{bezzina2015spmd} contains roughly 17,000,000 miles of data collected over almost 64,400 hours, including 17 TB of video. However, these datasets primarily focus on the capture of low-bandwidth data; the video streams of both datasets are compressed and downsampled to low frame rates.

Recently, high-quality computer vision-oriented datasets have been published. These include general driving datasets like Cityscapes~\cite{cordts2016cityscapes}, KITTI~\cite{geiger2013kitti} and BDD100K~\cite{yu2018bdd100k}, and traffic anomaly datasets like A3D~\cite{yao2019unsupervised}, DADA~\cite{fang2019dada}, and DoTA~\cite{yao2020when}.
Cityscapes contains 24,999 labelled images at 55 GB, while KITTI includes 7,481 images at 12 GB, in addition to 29 GB of point clouds and GPS and IMU data. BDD100K is one of the largest public driving datasets, having 100,000 HD video clips (1.8 TB) for over 1,100 driving hours in a variety of conditions. A3D, DADA and DoTA focus specifically on traffic anomalies. A3D contains 1,500 on-road accident clips with accident start and end times labelled. DADA releases 1,000 video clips with simulated driver eye-gaze. DoTA is comprised of 4,677 videos with spatial, temporal, and anomaly category annotations. Table~\ref{tab:dataset_summaries} summarizes some major driving video datasets.

\begin{table}[htbp]
    \caption{High-quality driving video datasets.}
    \centering
    \begin{tabular}{l|rlccc}
        \toprule
        Dataset & \multicolumn{2}{c}{\# frames} & Data size (GB) & Anomaly-focused & \# of anomlous videos\\
        \midrule
        KITTI & 7,481 & (15fps) & 12 & No&N/A\\
        Cityscapes & 24,999 & (17fps) & 55 & No&N/A\\
        BDD100K & 120,000,000 & (30fps) & $\sim$1,800 & No & N/A\\
        \midrule
        A3D & 128,174 & (10fps) & 15 & Yes & 1,500 \\
        DADA & 648,476 & (30fps) & 53 & Yes & 2,000 \\
        DoTA & 732,932 & (10fps) & 57 & Yes & 4,677\\
        \bottomrule
    \end{tabular}
    \label{tab:dataset_summaries}
\end{table}

Datasets like BDD100K and DoTA have significantly extended publicly available data access for deep-learning methods to use. However, anomaly-focused datasets are still relatively small; larger datasets like BDD100K contain very few EOIs with which to test self-driving algorithms. As a result, evaluation of the SBB required the creation of a combined dataset using BDD100K and DoTA video clips in order to have sufficient quantities of both normal and anomalous driving data. The SBB aims to address this problem by providing a method to collect high-value video data across an entire fleet of vehicles.
\section{Preliminaries} \label{sec:preliminary}
This work builds upon the Smart Black Box (SBB) intelligent event data recorder proposed in~\cite{yao2020sbb}. The original SBB design, data value estimation method, and its issues in the real world are reviewed.

\subsection{Smart Black Box Design} \label{sec:orig_sbb_design}
The SBB aims to record high-quality high-value data through value-driven data compression and prioritized data recording.
At each time step, one data frame is observed and collected. Based on event detectors, a scalar frame value $v_t\in[0,1]$ is computed for each frame. The data frame is then appended to a buffer, which caches seconds or minutes of data. The process of buffering data frames is managed by a deterministic Mealy machine (DMM) which uses the new data value, data similarity, and the current buffer size to determine when to end the current buffer and start a new one~\cite{yao2020sbb}. After the DMM terminates, local buffer optimization (LBO) is used to determine the optimal compression factor $d_t\in[0,1]$, called the LBO decision, for each frame in the buffer. A Gaussian data value filter can be applied over the buffered data to smooth the estimated data value. The buffered data is then compressed according to the LBO decisions and stored in long-term storage. After on-board storage is full, a priority-queue discards the lowest-value buffers to make space for higher-value buffers.

\subsection{SBB Value Estimation} \label{sec:orig_value_method}
The SBB was previously tested only in a simulation environment, The Open Source Racing Simulator (TORCS)~\cite{Espi2005TORCSTO}. Experiments done using TORCS in~\cite{yao2020sbb} classified each frame as either \textit{normal} ($\epsilon_1$) or as one of four pre-defined events of interest (EOIs): \textit{cutin}, \textit{hardbraking}, \textit{conflict}, or \textit{crash}, notated by $\epsilon_2,\epsilon_3,\epsilon_4, \mathrm{or\:} \epsilon_5$ respectively.
The value of each event is pre-computed using its event likelihood in Eq.~\ref{eq:orig_event_value}:
\begin{align} \label{eq:orig_event_value}
    v(\epsilon_j) = -\log_2(P(\epsilon_j))
\end{align}
where $P(\epsilon_j)$ is the likelihood of event $\epsilon_j$. These event values are then normalized over $[0,1]$ with $\max_jv(\epsilon_j)=1$. Frame value at time $t$, $v_t$, is then set according to
\begin{align}
    v_t = v(\epsilon(t))
\end{align}
where $\epsilon(t)$ is the event detected at time $t$.

This data value estimation method in works well in a simulation environment. However, it has two main drawbacks that affect its usability in the real-world. First, the method relies entirely on a set of pre-defined rules for EOI detection. In reality, the space of traffic EOIs is large and diverse, and capturing them purely using pre-defined rules is insufficient for real-world applications. Second, the detection of the four EOIs is not always possible given only dash camera data. In simulation, the EOIs are easily detectable by tracking the cars surrounding the ego vehicle. However, limiting the available sensing to a single front-facing camera makes the identification of these EOIs significantly more challenging.
In this paper, we apply an adjusted event classification system in Section~\ref{sec:frame_classification} and a new value estimation method in Section~\ref{sec:value_estimation}.
\section{Materials and Methods} \label{sec:materials_methods}
This section introduces a new event classification system to extend the previous SBB and defines updated data frame and buffer representations in Section~\ref{sec:data_cls}. Then, an updated SBB pipeline for real-world video data is presented in Section~\ref{sec:sbb_design}. Finally, methods for data value estimation using video anomaly detection and online action detection are discussed in Section~\ref{sec:value_estimation}.
\subsection{Data Classification and Representation}\label{sec:data_cls}

\subsubsection{Frame Classification} \label{sec:frame_classification}
As mentioned in Section~\ref{sec:orig_value_method}, the event classification system in \cite{yao2020sbb} does not straightforwardly apply to real-world applications. Instead, real-world datasets such as DoTA \cite{yao2020when} classify frames based on anomaly type and causation, e.g., a on-coming collision event. 
As such, we employ the classification system used in the DoTA dataset~\cite{yao2020when} which defines the eight traffic anomaly categories described in Table~\ref{tab:event_classes}. Each of these anomaly categories can be further specified as ego or non-ego events. Including the normal event class, this results in 17 total event classes. Online action detection aims to classify frames according to these event classes.

To realize generalized EOI detection, we also utilize a binary anomalous or normal classification. Video anomaly detection is used to solve this binary classification problem.

\begin{table}[htbp]
    \caption{Event Classes in the DoTA dataset ~\cite{yao2020when}. An anomaly label with "*" indicates an event where the ego car is not involved (\ie non-ego); otherwise the event is ego-involved.}
    \centering{}
    \begin{tabular}{ll|l}
        \toprule
        Name & ID & Description\\
        \midrule
        N & 0 & No anomaly\\
        ST & 1 & Collision with another vehicle which starts, stops, or is stationary\\
        AH & 2 & Collision with another vehicle moving ahead or waiting\\
        LA & 3 & Collision with another vehicle moving laterally in the same direction\\
        OC & 4 & Collision with another oncoming vehicle\\
        TC & 5 & Collision with another vehicle which turns into or crosses a road\\
        VP & 6 & Collision between vehicle and pedestrian\\
        VO & 7 & Collision with an obstacle in the roadway\\
        OO & 8 & Out-of-control and leaving the roadway to the left or right\\
        \bottomrule
    \end{tabular}\\
    \label{tab:event_classes}
\end{table}

\subsubsection{Data Frame Representation}
A \textit{data frame} is defined as all data, both observed and computed, associated with a single video frame. In this paper, we consider only data derived from camera input. This data includes:

\xhdr{Image}: The video frame captured by the camera. In this paper, we use RGB images at $1280\times720$ resolution.

\xhdr{Value}: The value of the frame $v_t\in[0,1]$. Value is calculated according to the value function defined in Section~\ref{sec:value_estimation}, and is used in the DMM as well as in buffer value computation.

\xhdr{Cost}: The normalized storage cost of the frame $c_t\in[0,1]$.

\xhdr{Anomaly score}: The anomaly score $s\in[0,1]$ of the frame generated using Video Anomaly Detection. More details can be found in Section \ref{sec:vad}.

\xhdr{Classification scores}: The output scores $\bm{o_t}=[o_{t,0},o_{t,1},\dots,o_{t,16}]$ for each event class in Table~\ref{tab:event_classes} from Online Action Detection. More details can be found in Section~\ref{sec:oad}.

\xhdr{Object data}: The tracking ID, object type, bounding box, and detector confidence of each object detected in the frame. Object data is used to support buffer tagging; details can be found in Section~\ref{sec:buffer_rep}.

\subsubsection{Frame Buffer Representation} \label{sec:buffer_rep}
A \textit{buffer} is a collection of  frames grouped by the DMM described in Section~\ref{sec:orig_sbb_design}. The buffer cost $C_k$ and value $V_k$ of the $k$th buffer are computed as
\begin{align}
V_k & = (1 + \lambda)^k\max_i(v_id_i)  \label{eq:buffer_value} \\
C_k & = \sum_i\hat{c_i} \label{eq:buffer_cost}
\end{align}
where $v_i$ is the value of the $i$th frame in the buffer, $d_i$ is its compression quality, and $\hat{c_i}$ is its post-compression storage cost. The $1+\lambda$ with $0<\lambda<<1$ is an aging factor used to slightly favor more recent buffers.

Additionally, \textit{buffer tags} are high-level descriptions of data buffers which enable buffer indexing and searching in downstream applications. These tags include:

\xhdr{Anomaly score}: The mean, max, and variance of the anomaly scores of the frames in the buffer.

\xhdr{Frame classifications}: A list of event classes $\epsilon$ for which there is a frame $f_t$ in the buffer where $o_{t,\epsilon}>\rho_\epsilon$ and $\rho_\epsilon$ is a user-defined threshold score for class $\epsilon$.

\xhdr{Objects}: The tracking ID, object type, and bounding boxes and detector confidences over time of each object in the buffer.
\subsection{Updated SBB Design}\label{sec:sbb_design}
\begin{figure}[hbtp]
    \centering
    \includegraphics[width=15cm]{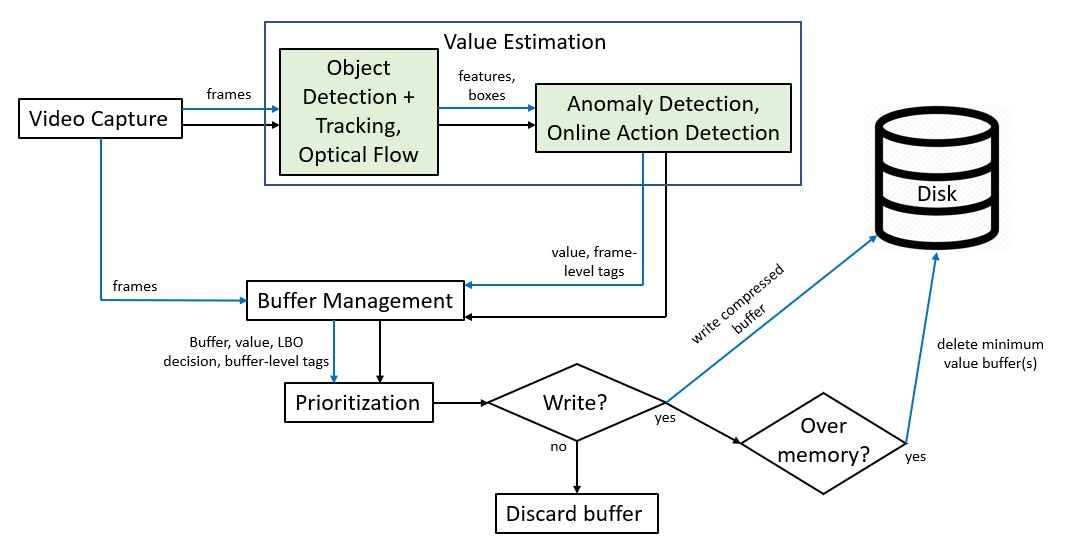}
    \caption{The updated SBB data recording pipeline. Green blocks denote new modules that were not present in the original SBB~\cite{yao2020sbb}. Black arrows indicate logic flow and blue arrows indicate data flow.}
    \label{fig:mp_sbb}
\end{figure}

The updated SBB is separated into four processes running in parallel: video capture, buffer management, value estimation, and prioritization. Figure~\ref{fig:mp_sbb} describes the updated SBB pipeline. Our code is available in \href{https://github.com/rzf16/sbb2_algs}{https://github.com/rzf16/sbb2\_algs}.

\xhdr{Video Capture}  reads video input and publishes each video frame to value estimation and buffer management. This module remains unchanged from the original SBB.

\xhdr{Value Estimation}  assigns a value $v_t\in[0,1]$ for each frame to be used in buffer management and storage prioritization. The value estimation module first executes object detection, object tracking~\cite{wojke2017simple}, and optical flow estimation~\cite{ilg2016flownet}. The outputs are then used in Video Anomaly Detection and Online Action Detection, which are used to compute the value. Details on this calculation can be found in Section~\ref{sec:value_estimation}.
The value estimation method used differs significantly from~\cite{yao2020sbb}. In~\cite{yao2020sbb}, perfect EOI detection using pre-defined rules was assumed, and data value was computed based on detected EOIs. In this paper, we instead use video anomaly detection and action detection methods for generalized EOI detection and calculate data value based on their output scores. More details on value estimation of the original SBB can be found in Section~\ref{sec:orig_value_method}. More details on our updated value estimation can be found in Section~\ref{sec:value_estimation}.

\xhdr{Buffer Management} \label{sec:buffer_management} groups frames into buffers using the DMM from \cite{yao2020sbb} after receiving each frame from Video Capture and its corresponding value from Value Estimation. The similarity of a data frame to the current buffer is computed as the percentage of object tracking IDs in the frame which have already appeared in the buffer. With $A$ being the set of tracking IDs in the frame and $B$ being the set of tracking IDs which have appeared in the buffer, we compute the similarity $\xi_t=\frac{|A \cap B|}{|A|}$. Once the DMM terminates, LBO solves an optimization problem over the output buffer to determine the compression quality of each frame.

According to \cite{yao2020sbb}, a decoupled LBO strategy can optimize the compression quality of a single frame independent of all other frames in the buffer. Given constant $\frac{\eta}{\zeta}$, the uncoupled LBO objective function for frame $f_t$ is:
\begin{align} \label{eq:lbo_obj}
\begin{split}
    \min_{d_t}c_t\phi(d_t) - \frac{\eta}{\zeta}\hat{v_t}d_t\\
    \mathrm{subject\:to\:} d_t\in[0,1]
\end{split}
\end{align}
where $\eta,\zeta\geq0$ are weighting parameters and $\phi(d_t)$ maps from the compression quality to the compression ratio. Note that $\phi(d_t)$ increases monotonically over $d_t\in[0,1]$. In this paper, we use the $\phi$ function of JPEG compression on real-world driving data following~\cite{yao2018smart}. 
Throughout the paper, the values $\eta=0.9$ and $\eta=1.7$ are used based on~\cite{yao2020sbb}. These parameters were assigned to maximize value-per-memory (VPM) of the recorded data. Further details on $\eta$ and $\zeta$ parameter selection can be found in~\cite{yao2020sbb}.

DMM and LBO functionality remain the same as~\cite{yao2020sbb}. However, the data similarity metric is adjusted to match our focus on dash camera data. In the previous SBB, data similarity was computed using the odometry of the host and surrounding vehicles. However, a single front-facing camera cannot capture sufficient information to use this approach. As such, we compute data similarity using detected objects in the frame, as mentioned above.

\xhdr{Prioritization} maintains a buffer priority heap in order to retain high-value buffers and delete low-value buffers as the memory capacity is reached. Buffer value $V_k$ and cost $C_k$ of the $k$th buffer are computed according to Eq.~\ref{eq:buffer_value} and Eq.~\ref{eq:buffer_cost} respectively. A binary min-heap is constructed to store buffers based on $V_k$ following~\cite{yao2020sbb}. This module is also unchanged from~\cite{yao2020sbb}.

\subsection{Value Estimation Method} \label{sec:value_estimation}

This section introduces the data value estimation method used by the DMM module to group buffers and decide the optimal compression factors. 
Similar to~\cite{yao2018smart,yao2020sbb}, we define the value of a data frame as a measure of data anomaly.  The data value is determined by: 1) The anomaly score estimated by a video anomaly detection (VAD) module; 2) The anomaly category detected by an online action detection (OAD) module.

\subsubsection{Video Anomaly Detection (VAD)} \label{sec:vad}
A VAD algorithm takes observed image frames and predicts an anomaly score for each frame as a description of the degree of abnormality of that frame. Existing VAD algorithms can be categorized as frame-level VAD and object-level VAD. A frame-level VAD algorithm reconstructs or predicts image frames (e.g., in RGB or grayscale) and computes the L2 error of reconstruction or prediction as the anomaly score~\cite{hasan2016learning,chong2017abnormal,liu2018future}. An object-level algorithm, on the other hand, predicts object appearance and/or motions and computes the anomaly score based on prediction error~\cite{ionescu2019object,morais2019learning} or consistency~\cite{yao2019unsupervised,yao2020when}.

In this paper, we run an off-the-shelf VAD algorithm to estimate
an anomaly score $s_{t}$ of a frame $f_t$ and use it to inform our value estimation. 
To be specific, we trained the TAD algorithm in~\cite{yao2019unsupervised} using the Detection of Traffic Anomaly (DoTA) dataset following~\cite{yao2020when} and applied it to our data value estimation module.

\subsubsection{Online Action Detection (OAD)} \label{sec:oad}

While the anomaly score from VAD provides information about the probability an anomaly occurs in a frame, it does not assess anomaly category which is important information for determining data value in long-term driving according to~\cite{yao2020sbb}.
Categorizing anomalous events is essential to the SBB design since it allows the SBB to prioritize high value categories when the storage limit is encountered, and it allows the SBB to focus on specific event types per a user's request. 

In this paper, we implement an off-the-shelf OAD algorithm to obtain a confidence score vector $o_{t}$ for a frame $f_t$, which is then combined with the anomaly score $s_{t}$ to estimate the data value. To be specific, We trained an OAD algorithm called the temporal recurrent network (TRN)~\cite{xu2019temporal} using the DoTA dataset~\cite{yao2020when}. The TRN outputs a 17-D vector $o_{t}=[P_{t}(\epsilon_{1}),P_{t}(\epsilon_{2}),\dots,P_{t}(\epsilon_{17})]$ for each frame with $\sum_{j=0}^{16}P_{t}(\epsilon_{j})=1$ which represents the confidence score that a frame belongs to each class.

\subsubsection{Hybrid Value}
A hybrid value estimation method is proposed which sums VAD and OAD scores according to 
\begin{align} \label{eq:hybrid_value}
v = \max(1,\:\alpha s + \beta\sum_{i=1}^{16}w_io_i)
\end{align}
where $w_i$ is the information measure of class $i$ and $\alpha$, $\beta$ are weighting parameters in $[0,1]$. Because $o_i$ estimates the probability that a frame is of class $i$, the weighted sum over $\bm{o}$ is equivalent to the expected information measure of the frame. By using the information measure of each anomaly class, anomaly types of higher rarity are assigned a higher value. Note that class 0, the normal class, is not included in the computation. This is equivalent to $w_0=0$. Throughout this paper, we use $\alpha=\beta=1$ for simplicity. 

The information measures $w_i$ for each class are calculated using the class likelihoods in the DoTA dataset found in Table~\ref{tab:class_probs_vals}. The information measure $w_i$ for class $i$ is calculated in Eq.~\eqref{eq:info_measure}. Values are normalized to $[0,1]$ by dividing by the maximum information measure.

\begin{align}\label{eq:info_measure}
    w_i=-\log_2(P(\mathrm{class}=i))
\end{align}

\begin{table}[htbp]
    \caption{DoTA Anomaly Class Probabilities and Values}
    \centering
    \resizebox{1.0 \textwidth}{!} {
    \begin{tabular}{c|ccccccccccccccccc}
        \toprule
        & ST & AH & LA & OC & TC & VP & VO & OO & ST* & AH* & LA* & OC* & TC* & VP* & VO* & OO*\\
        \midrule
        Likelihood & 0.011 & 0.057 & 0.054 & 0.023 & 0.163 & 0.012 & 0.010 & 0.089 & 0.010 & 0.091 & 0.104 & 0.081 & 0.207 & 0.010 & 0.011 & 0.070\\
        \midrule
        \makecell{Normalized \\info. measure} & 0.977 & 0.635 & 0.633 & 0.816 & 0.395 & 0.957 & 0.995 & 0.525 & 1.0 & 0.521 & 0.491 & 0.546 & 0.342 & 1.0 & 0.990 & 0.576\\
        \bottomrule 
    \end{tabular}}
    \label{tab:class_probs_vals}
\end{table}

\section{Experiments} \label{sec:experiments}
In this section, we conduct SBB data collection experiments on a specifically designed large-scale real-world video dataset and present the results. We discuss storage requirements of SBB-compressed data to showcase its preservation of valuable data. We then compare our SBB prioritized data recording with a FIFO queue data recording strategy. We examine SBB results on each anomaly class. Finally, we evaluate the performance of the VAD-OAD hybrid value method with several parameter combinations.

\subsection{Dataset}
The SBB is designed for high-bandwidth data collection in long-term driving where on-board storage is limited. Therefore, SBB performance evaluation requires a large, high-quality video dataset which contains both normal driving data as well as events of interest (EOIs). To our best knowledge, there is currently no single dataset that satisfies all these requirements. The BDD100K dataset~\cite{yu2018bdd100k} is one of the largest high-quality driving video datasets and contains $100,000$ video clips covering about $\sim1,100$ driving hours. The DoTA dataset~\cite{yao2020when} is the largest and newest high-quality video dataset for traffic anomalies and contains $4,677$ anomalous video clips. We combined the $10,000$ validation videos in the BDD100K dataset and randomly sampled and interspersed $500$ anomalous video clips from the DoTA dataset, resulting in a large testing video with $\sim4,000,000$ frames at $10$ FPS. The frames from BDD100K were compressed using OpenCV with JPEG quality 85 in order to eliminate the difference in image size between BDD100K and DoTA. This combined dataset contains over 100 hours of driving video at $1280\times720$ resolution with ~0.5\% of frames being anomalous, meeting our requirements of a large, high-quality, and mostly non-anomalous dataset. Note that the ST* anomaly class was not included in this combined dataset, as its rarity in the DoTA dataset led to no ST* clips being sampled.

\subsection{Results}
Experimental results for the SBB are presented and discussed below. In general, high storage size and decisions are desirable for anomalous data, while the opposite is true for normal data.

\begin{figure}[h]
    \centering
    \subfloat[A normal driving frame]{
        \includegraphics[width=7.5cm]{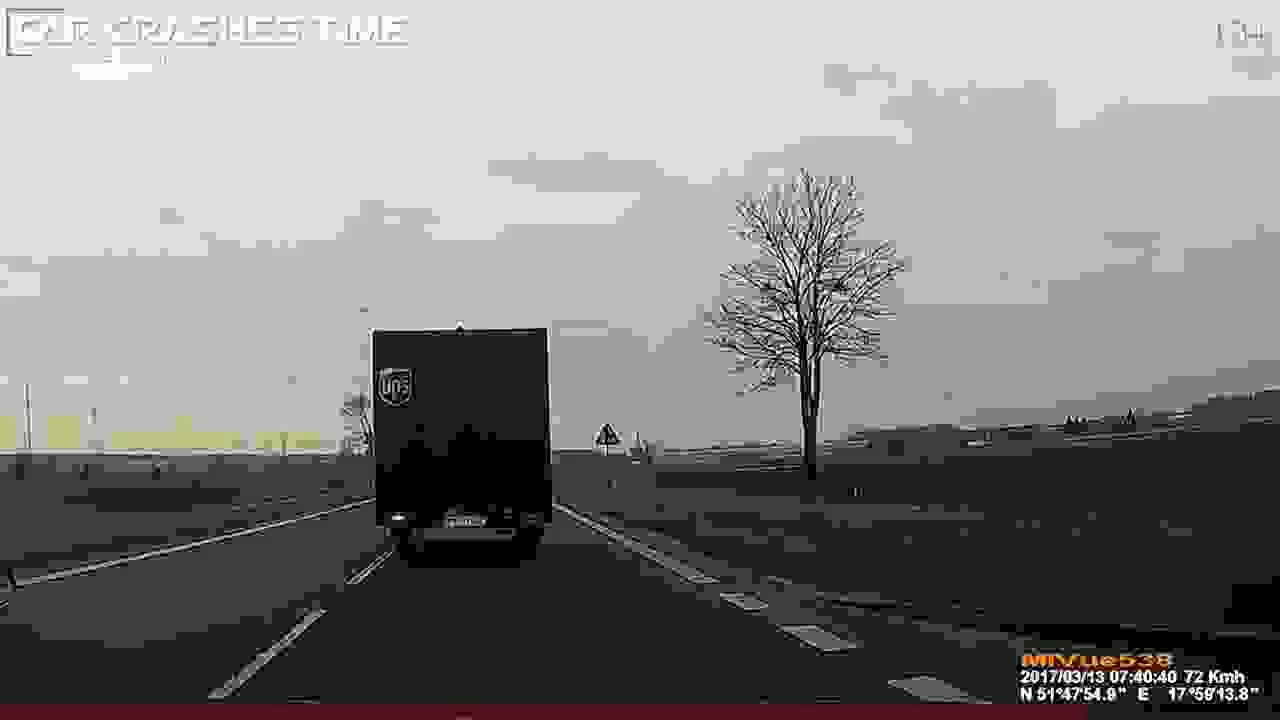}
    }
    \subfloat[A vehicle-object collision (VO) event]{
        \includegraphics[width=7.5cm]{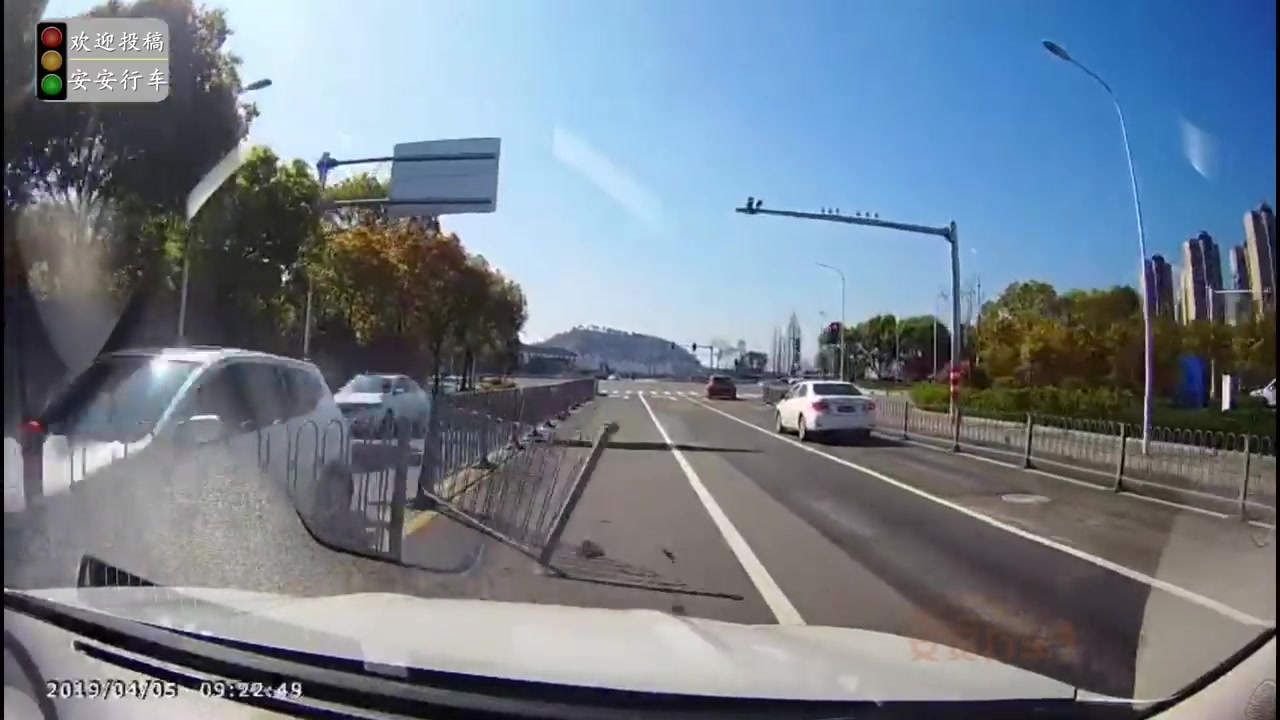}
    }\\
    \subfloat[Precursor of the TC event]{
        \includegraphics[width=7.5cm]{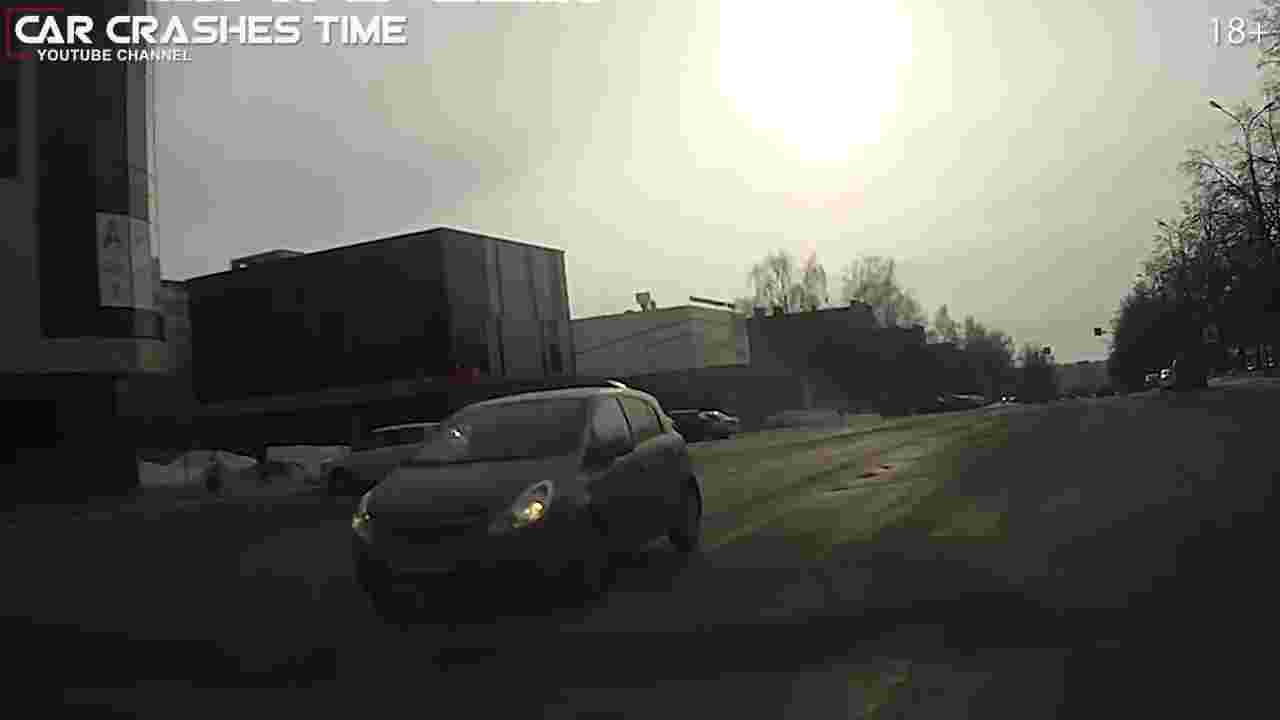}
    }
    \subfloat[An ego turning collision (TC) event]{
        \includegraphics[width=7.5cm]{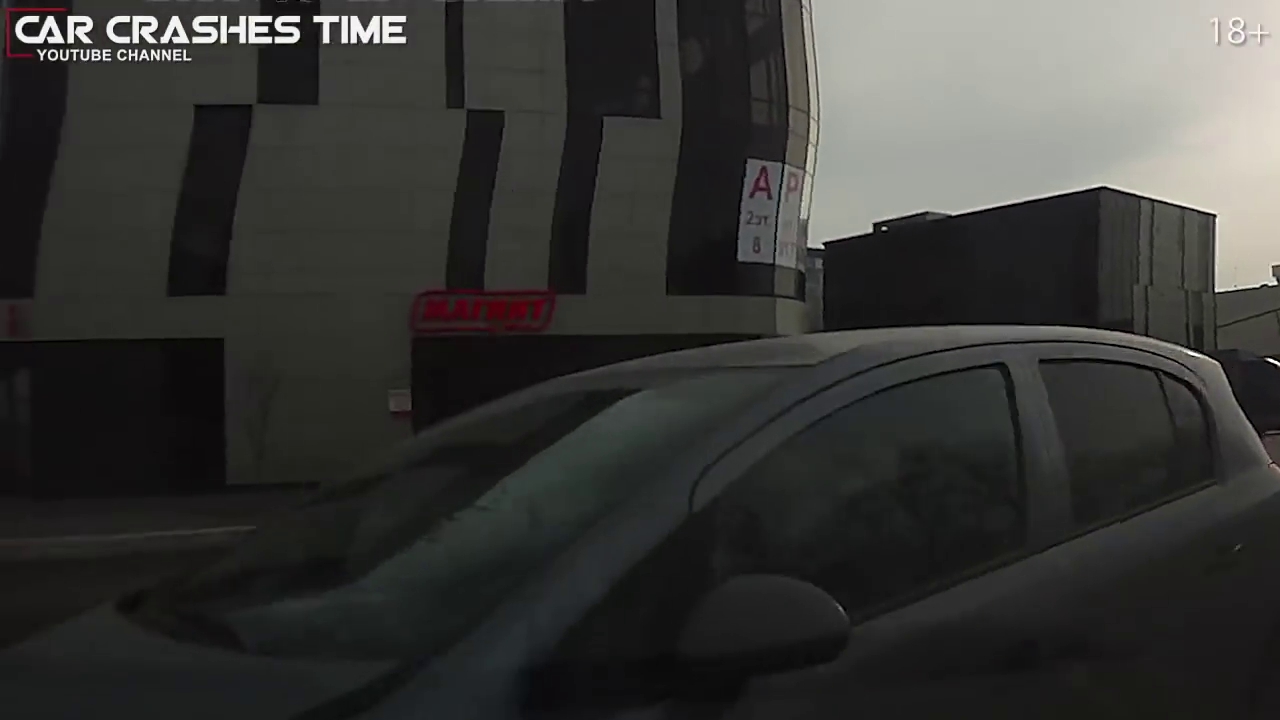}
    }
    \caption{Compressed normal frames (left) and preserved anomalies (right)}
    \label{fig:compression_comp}
\end{figure}

\begin{figure}[h]
    \centering
    \subfloat[A non-ego out-of-control (OO*) event where a windshield wiper is partially blocking the anomaly]{
        \includegraphics[width=7.5cm]{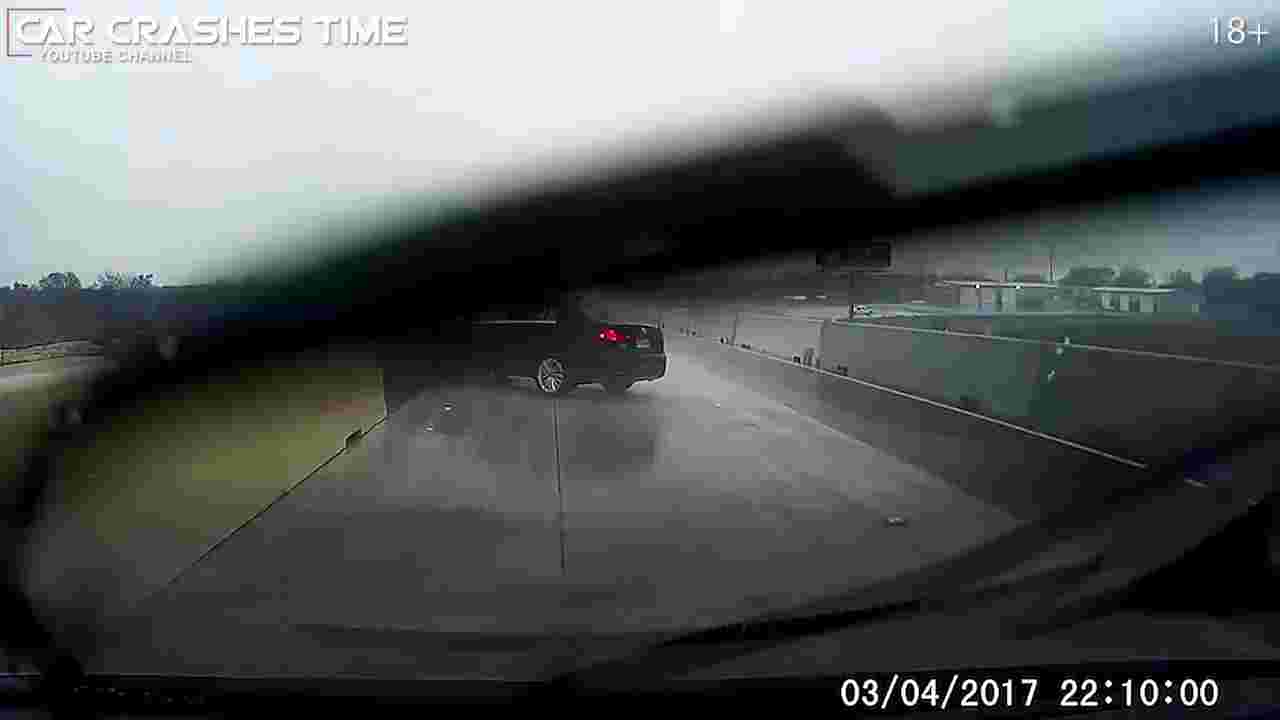}
    }
    \subfloat[A non-ego road crossing collision (TC*) where one vehicle is partially occluded]{
        \includegraphics[width=7.5cm]{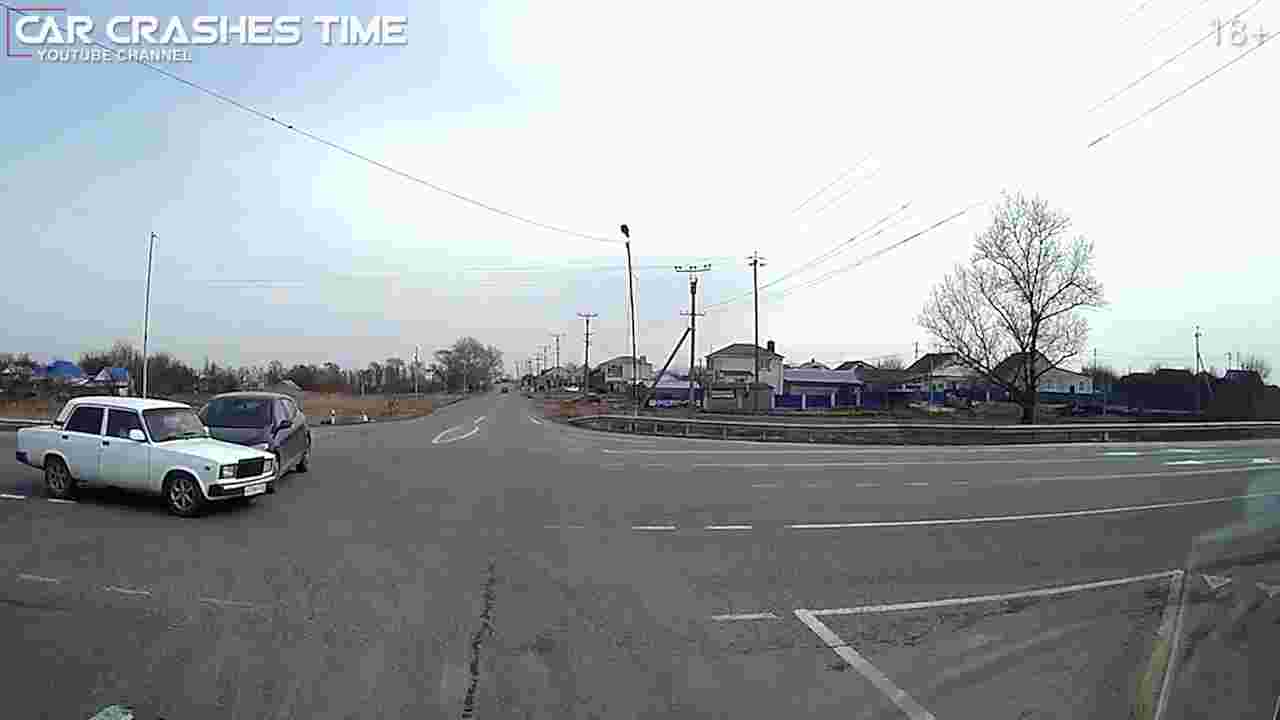}
    }
    \caption{Compressed anomaly failure cases}
    \label{fig:compressed_anoms}
\end{figure}

\xhdr{SBB Data Compression.} SBB data compression statistics with no memory limit are presented in Table~\ref{tab:sbb_dota_result_1}. It can be seen that the storage cost of normal frames is significantly reduced (427.20 GB to 65.53 GB, 85\%) by the SBB. This leads to a 24.4\% increase in the ratio of anomalous data storage to normal data storage.
Both the average (\textbf{avg.}) and median (\textbf{med.}) compression factor decisions of the SBB are higher for the anomalous frames, indicating that the SBB is able to identify and preserve anomalous frames over normal ones. Figure~\ref{fig:compression_comp} displays normal frames which were highly compressed by the SBB along with preserved anomalous frames; Figure~\ref{fig:compressed_anoms} shows two failure cases where anomalous frames were mistakenly compressed. Both of these failures showcase a lack of robustness against cases where anomalous objects are occluded.

However, the decision difference between normal and anomaly frames is not as significant compared to the simulation experiment in \cite{yao2020sbb} due to the fact that the anomalous event detection in simulation was $100\%$ accurate while VAD on real-world data is far from perfect. Moreover, the median decision for a normal frame is significantly lower than the mean, indicating that there are outlier normal frames with unusually high value scores. The standard deviation (\textbf{std.}) of anomalous frames is significantly larger than that in the simulation experiment ($0.26$ vs $\sim0.02$), showing how inaccurate VAD and OAD reduces the SBB's efficiency on real-world data.

The limitations of VAD and OAD are further shown by evaluating performance of the SBB given ground-truth labels as VAD and OAD scores. The anomalous-to-normal memory  ratio increases by 1610\%, driven by the massive difference in decisions between normal and anomalous frames. This upper-bound performance of the SBB indicates that as anomaly detection techniques continue to improve, the performance of the SBB will improve as well.

\begin{table}[htbp]
    \caption{Raw, SBB-compressed, and ground-truth (GT) SBB-compressed data statistics on the BDD100K+DoTA dataset.}
    \centering
    \begin{tabular}{l|l|ccc}
        \toprule
         &  & Normal & Anomaly & Anomaly Ratio\\
        \midrule
        \multirow{2}{*}{Raw Data} & \# of frames & 3,967,977 & 16,768 &\\
         & size (GB) & 427.20 & 1.76 & 0.41\%\\
        \midrule
        \multirow{2}{*}{SBB w/ VAD+OAD}  & size (GB) & 65.53 & 0.33 & 0.51\% \\
        & \textbf{avg.} $d_{i}$ & 0.51 & 0.58 &\\
        & \textbf{med.} $d_{i}$ & 0.55 & 0.63 &\\
        & \textbf{std.} $d_{i}$ & 0.24 & 0.26 &\\
        \midrule
        \multirow{2}{*}{SBB w/ GT VAD+OAD}  & size (GB) & 11.05 & 0.73 & 6.60\%\\
        & \textbf{avg.} $d_{i}$ & 0.00 & 0.92 &\\
        & \textbf{med.} $d_{i}$ & 0.00 & 0.92 &\\
        & \textbf{std.} $d_{i}$ & 0.03 & 0.00 &\\
         \bottomrule
    \end{tabular}
    \label{tab:sbb_dota_result_1}
\end{table}

\xhdr{Priority Queue vs. FIFO.} Table~\ref{tab:sbb_dota_fifo} compares the recorded frames of a prioritized recording system against that of a FIFO queue at memory limits of $M=3.125$ GB, $6.25$ GB, $12.5$ GB and $25$ GB. These values represent a non-trivial amount of data to upload (depending on internet connection quality) assuming continuous internet access is not available. In all scenarios,
the prioritized recording saved fewer normal frames and more
anomalous frames than with the FIFO strategy. We also note that while the anomaly ratio stays roughly the same in each memory limit for the FIFO queue, the ratio increases at each level for the priority queue. The prioritization strategy of the SBB removes $\sim95\%$ of the normal frames while still recording $\sim10\%$ anomalous frames at $M=3.125$ GB. Compared to the FIFO queue, the anomalous-to-normal count ratio of SBB-recorded data is $\sim25\%$ to $\sim100\%$ higher.

\begin{table}[htbp]
    \caption{Comparison of Prioritized Recording and FIFO.}
    \centering{}%
    \begin{tabular}{l|l|ccc}
        \toprule
        $M$&  & Normal$\downarrow$ & Anomaly$\uparrow$ & Anomaly Ratio$\uparrow$\\
        \midrule
        \multirow{2}{*}{25 GB} &  FIFO & 1,739,855 & 7851 & 0.45\%\\
        & \textbf{Priority} & \textbf{1,487,570} & \textbf{8545} & \textbf{0.57\%}\\
        \midrule
        \multirow{2}{*}{12.5 GB} & FIFO & 889,679 & 4335 & 0.49\%\\
        & \textbf{Priority} & \textbf{734,625} & \textbf{5,154} & \textbf{0.70\%}\\
        \midrule
        \multirow{2}{*}{6.25 GB} & FIFO & 437,673 & 2029 & 0.46\%\\
        & \textbf{Priority} & \textbf{364,666} & \textbf{2,898} & \textbf{0.79\%}\\
        \midrule
        \multirow{2}{*}{3.125 GB} & FIFO & 207,951 & 962 & 0.46\%\\
        & \textbf{Priority} & \textbf{183,951} & \textbf{1,706} & \textbf{0.93\%}\\
        \bottomrule 
    \end{tabular}
    \label{tab:sbb_dota_fifo}
\end{table}

\begin{figure}[h]
    \centering
    \subfloat[ST]{
        \includegraphics[width=3cm]{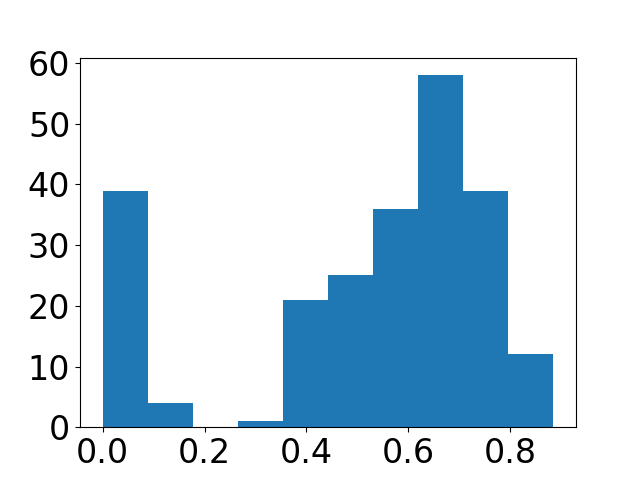}
    }
    \subfloat[AH]{
        \includegraphics[width=3cm]{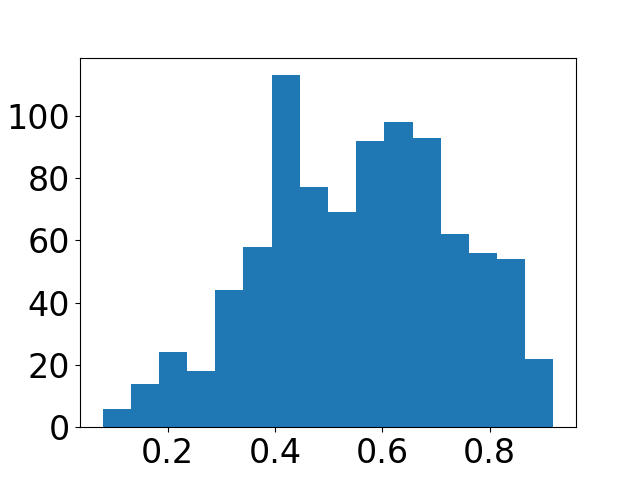}
    }
    \subfloat[LA]{
        \includegraphics[width=3cm]{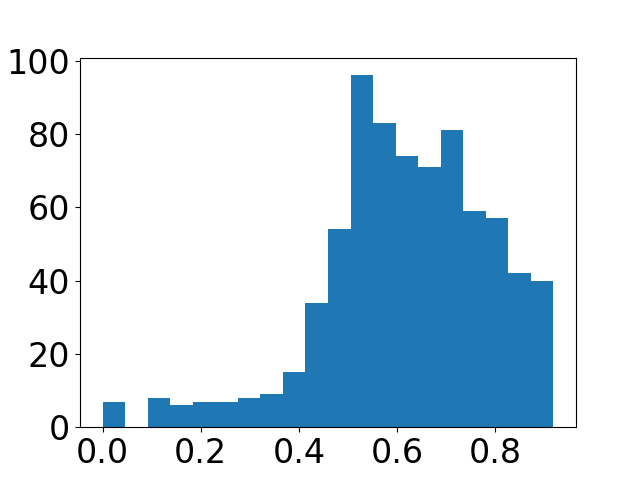}
    }
    \subfloat[OC]{
        \includegraphics[width=3cm]{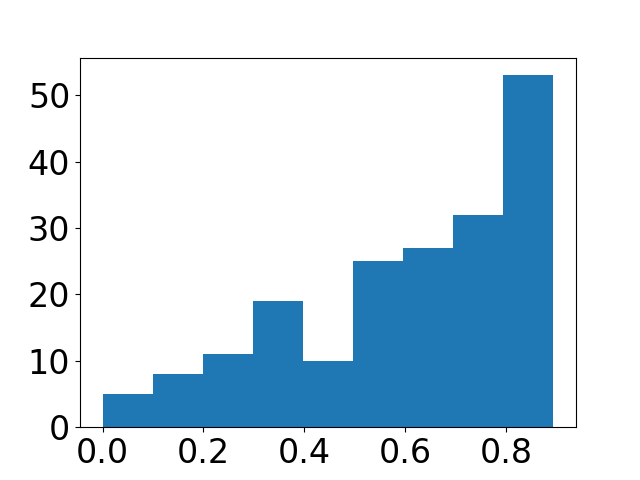}
    }\\
    \subfloat[TC]{
        \includegraphics[width=3cm]{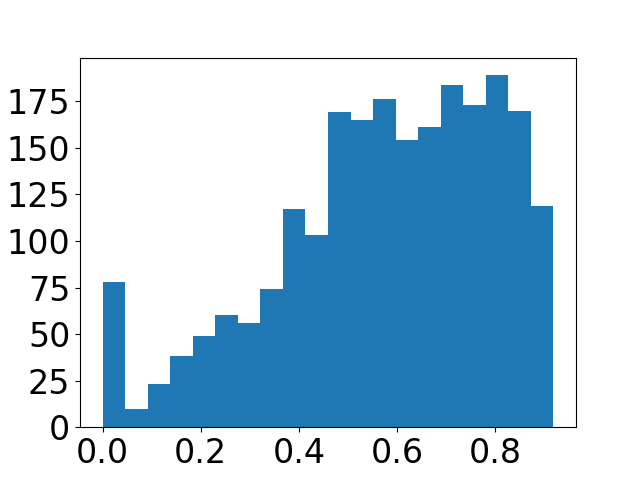}
    }
    \subfloat[VP]{
        \includegraphics[width=3cm]{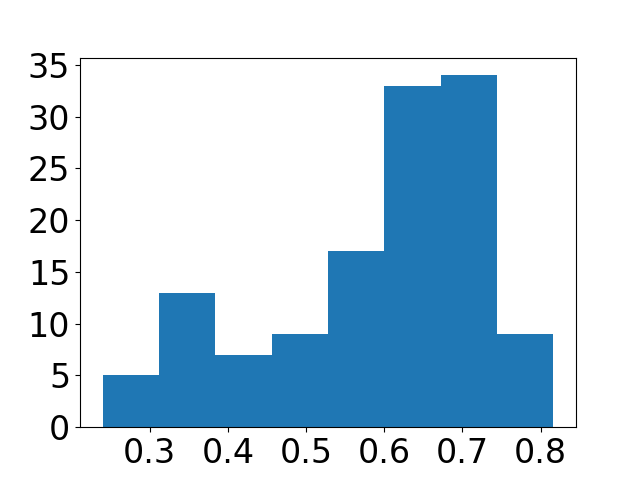}
    }
    \subfloat[VO]{
        \includegraphics[width=3cm]{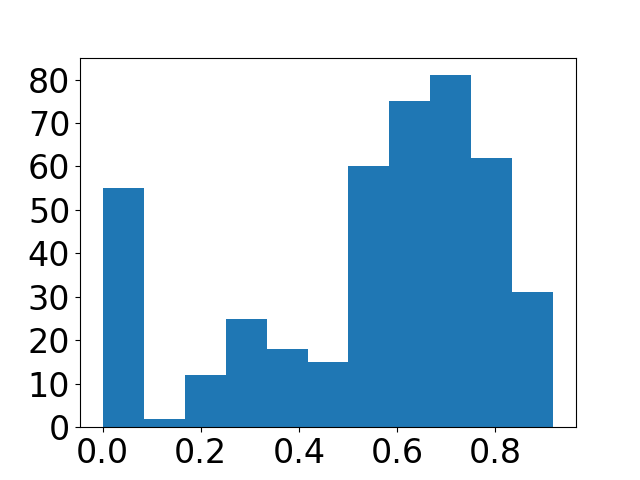}
    }
    \subfloat[OO]{
        \includegraphics[width=3cm]{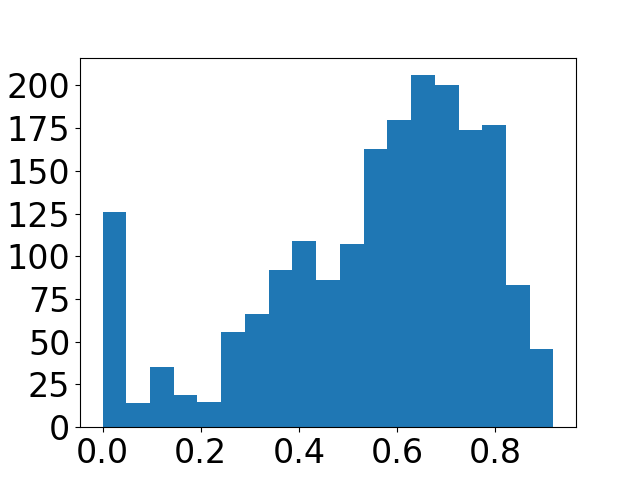}
    }\\
    \subfloat[ST*]{
        \includegraphics[width=3cm]{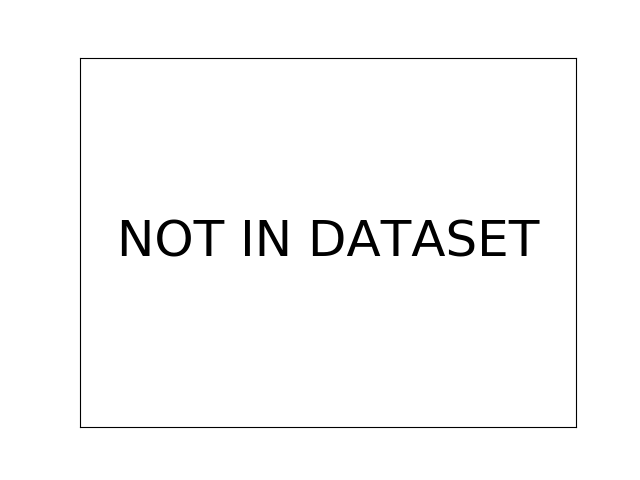}
    }
    \subfloat[AH*]{
        \includegraphics[width=3cm]{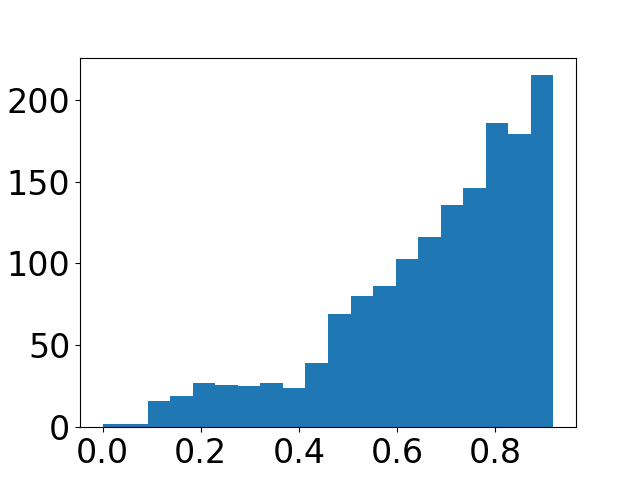}
    }
    \subfloat[LA*]{
        \includegraphics[width=3cm]{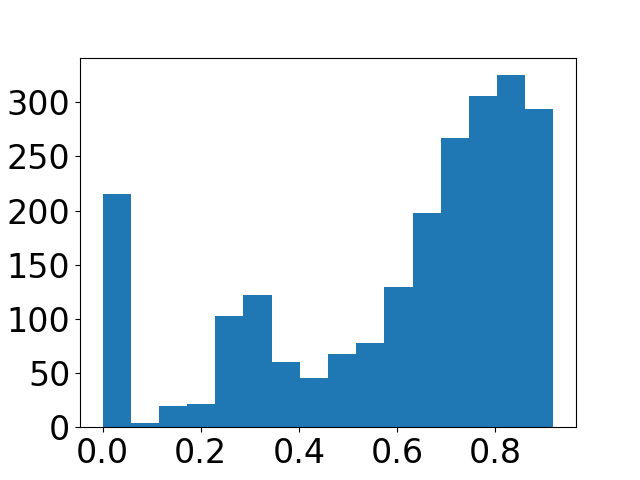}
    }
    \subfloat[OC*]{
        \includegraphics[width=3cm]{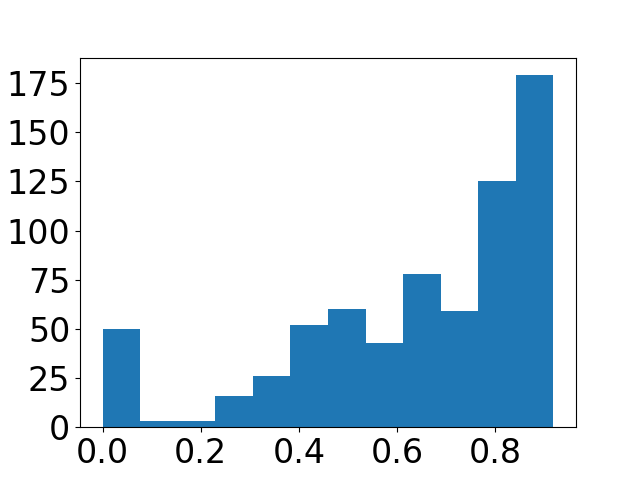}
    }\\
    \subfloat[TC*]{
        \includegraphics[width=3cm]{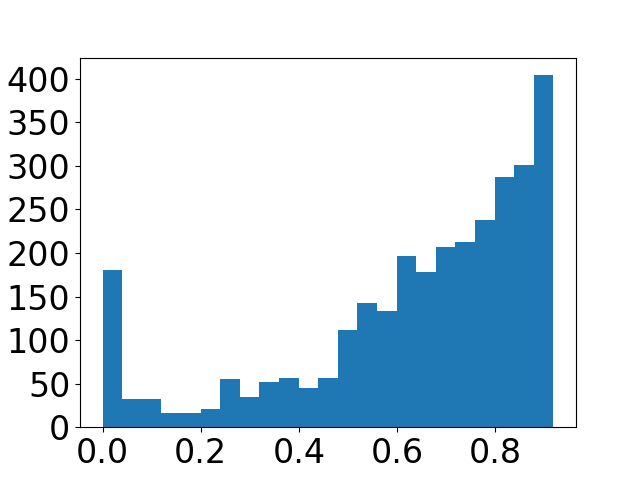}
    }
    \subfloat[VP*]{
        \includegraphics[width=3cm]{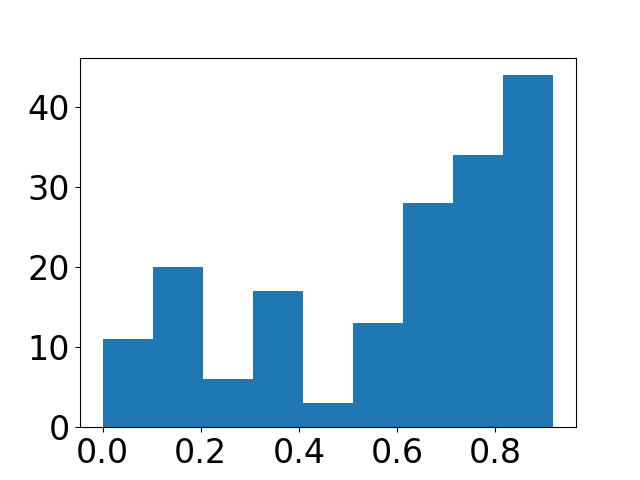}
    }
    \subfloat[VO*]{
        \includegraphics[width=3cm]{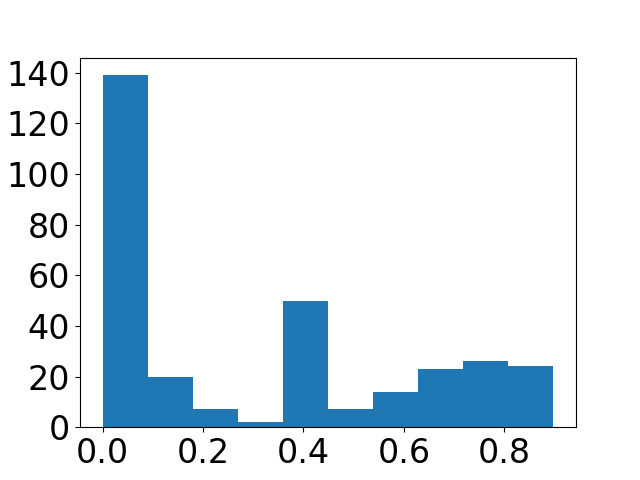}
    }
    \subfloat[OO*]{
        \includegraphics[width=3cm]{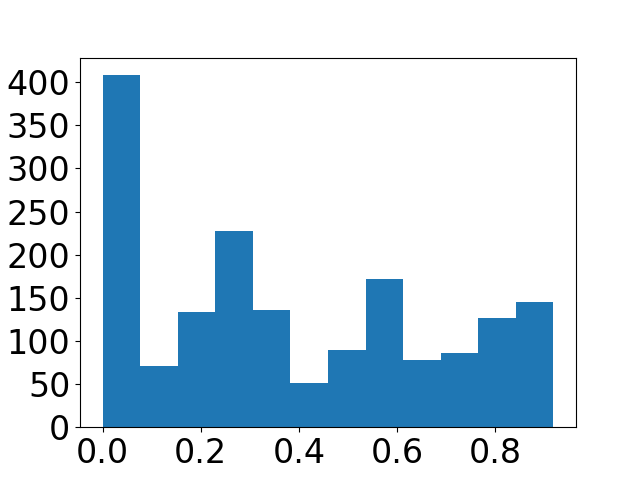}
    }
    \caption{Decision histograms per anomaly class\protect\footnotemark. The $x$-axis represents decision $d_t\in[0,1]$, and the $y$-axis represents the frame count per bin.}
    \label{fig:per_class_decision_dists}
\end{figure}

\xhdr{Performance Per Anomaly Class.}
Figure~\ref{fig:per_class_decision_dists} displays the decision histograms for each anomaly class. The performance of the SBB varies heavily depending on the anomaly category. For example, the decision distribution of class OC indicates very good detection of this anomaly. In OC, an ego-vehicle collision with an oncoming vehicle, the anomalous object (the oncoming vehicle) is almost always both near the camera and largely unoccluded. However, ST, VO, LA*, VO*, and OO* have notably poor performance. ST is an extremely difficult case for OAD due to its visual similarity to AH and LA anomalies, resulting in lower OAD confidence that an anomaly has occurred. VO and VO* involve vehicles hitting obstacles in the roadway. In some scenarios, such as hitting a traffic cone or a fire hydrant, the obstacle may be blocked from view by the anomalous vehicle in a non-ego incident or outside the camera's field of view in an ego-incident. LA* often involves vehicles slowly moving closer together, making the collision relatively subtle. OO*, a non-ego vehicle leaving the roadway, can be challenging to detect simply due to the distance at which the anomaly occurs. 

\footnotetext{\textbf{ST:} collision with a starting, stopping, or stationary vehicle; \textbf{AH:} ahead collision; \textbf{LA:} lateral collision; \textbf{OC:} oncoming collision; \textbf{TC:} turning or crossing collision; \textbf{VP:} vehicle-pedestrian collision; \textbf{VO:} vehicle-obstacle collision; \textbf{OO:} out-of-control leaving roadway; \textbf{*} indicates non-ego}

\xhdr{Value Estimation Method Comparison.}
Table~\ref{tab:value_methods} compares decision statistics for hybrid value estimation with several parameter combinations. We note that the VAD-only method generates the largest decision difference in normal and anomalous frames; we suspect this to be a result of OAD's inability to consistently differentiate between anomalous and normal frames. Readers are directed to~\cite{yao2020when} for an in-depth discussion on the poor performance of OAD algorithms.

\begin{figure}[h]
    \centering
    \includegraphics[width=14cm]{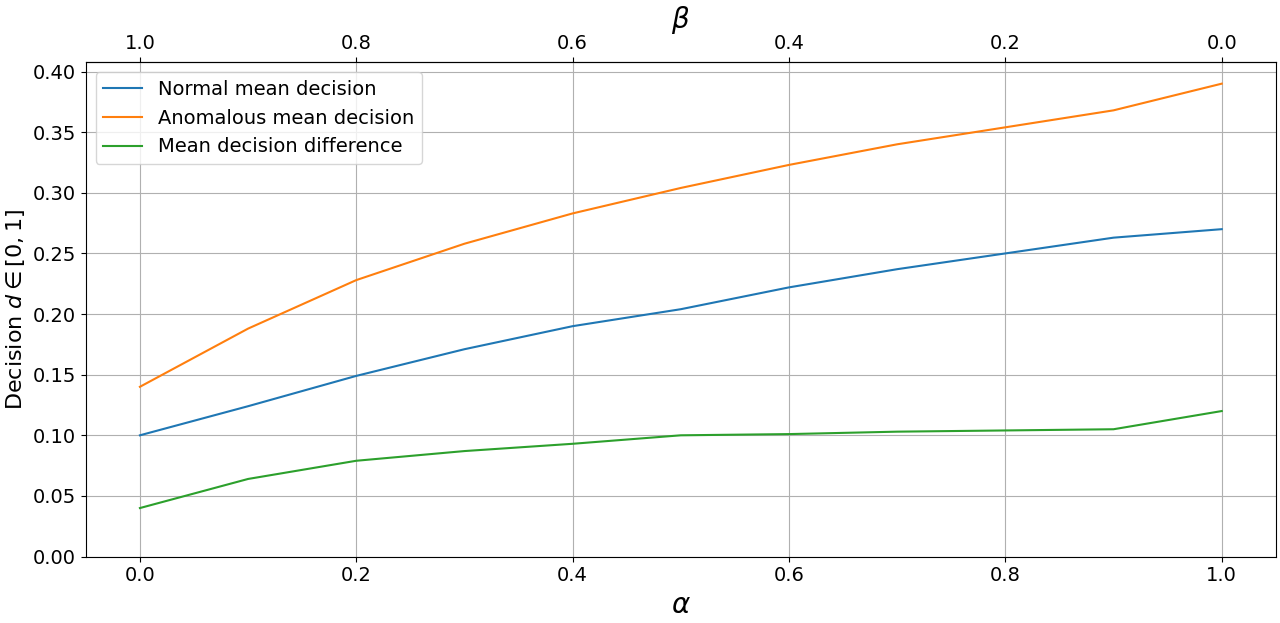}
    \caption{Mean normal and anomalous decisions for $\alpha+\beta=1$.}
    \label{fig:hybrid_params}
\end{figure}

For applications which value general EOIs, VAD-only value estimation ($\alpha=1$, $\beta=0$) has the greatest ability to distinguish normal and anomalous data. However, users interested in specific EOIs may opt to use hybrid  value in order to incorporate the EOI classification offered by OAD. In terms of hybrid value parameters, Table~\ref{tab:value_methods} shows that lower weights result in higher decision differences. However, in situations where retaining data quality is critical, higher $\alpha$ and $\beta$ values may be used to achieve higher overall decision quality. Additionally, the higher decision differences as $\alpha$ increases shown in Figure~\ref{fig:hybrid_params} indicate once again that VAD contributes more to the differentiation of normal and anomalous frames than does OAD.

\begin{table}[htbp]
    \caption{Compression quality decisions for hybrid value estimation.}
    \centering
    \begin{tabular}{c|cc|l|cc}
        \toprule
        Value Estimation & $\alpha$ & $\beta$ & & Normal & Anomaly\\
        \midrule
        \multirow{3}{*}{VAD Only}& \multirow{3}{*}{1.0} & \multirow{3}{*}{0.0} &
        \textbf{avg.} $d_{i}$ & 0.27 & 0.38\\
        & & & \textbf{med.} $d_{i}$ & 0.15 & 0.38\\
        & & & \textbf{std.} $d_{i}$ & 0.30 & 0.34\\
        \midrule
        \multirow{3}{*}{OAD Only}& \multirow{3}{*}{0.0} & \multirow{3}{*}{1.0} &
        \textbf{avg.} $d_{i}$ & 0.10 & 0.14\\
        & & & \textbf{med.} $d_{i}$ & 0.0 & 0.07\\
        & & & \textbf{std.} $d_{i}$ & 0.15 & 0.17\\
        \midrule
        \multirow{12}{*}{Hybrid} & \multirow{3}{*}{1.0} & \multirow{3}{*}{1.0} & \textbf{avg.} $d_{i}$ & 0.51 & 0.58\\
        & & & \textbf{med.} $d_{i}$ & 0.55 & 0.63\\
        & & & \textbf{std.} $d_{i}$ & 0.24 & 0.26\\
        \cmidrule{2-6}
        & \multirow{3}{*}{0.9} & \multirow{3}{*}{0.1} & \textbf{avg.} $d_{i}$ & 0.26 & 0.37\\
        & & & \textbf{med.} $d_{i}$ & 0.14 & 0.36\\
        & & & \textbf{std.} $d_{i}$ & 0.29 & 0.33\\
        
        \cmidrule{2-6}
        & \multirow{3}{*}{0.5} & \multirow{3}{*}{0.5} & \textbf{avg.} $d_{i}$ & 0.20 & 0.30\\
        & & & \textbf{med.} $d_{i}$ & 0.10 & 0.27\\
        & & & \textbf{std.} $d_{i}$ & 0.24 & 0.28\\
        \cmidrule{2-6}
        & \multirow{3}{*}{0.1} & \multirow{3}{*}{0.9} & \textbf{avg.} $d_{i}$ & 0.12 & 0.19\\
        & & & \textbf{med.} $d_{i}$ & 0.02 & 0.16\\
        & & & \textbf{std.} $d_{i}$ & 0.16 & 0.19\\
        \bottomrule
    \end{tabular}
    \label{tab:value_methods}
\end{table}
\section{Discussion}
While our results indicate that the SBB improves the recording of anomalous data, the massive difference between the ground truth and actual performance of the SBB indicate a need for improved video-based anomaly detection and action detection. Furthermore, a key limitation of our study is the extraction of data value purely from a single video stream. Most autonomous vehicles feature huge sensor suites, including multiple cameras, radar, lidar, CAN data, \etc. Future research in intelligent event data recorders may tap into this wealth of sensor data to more effectively detect EOIs and assign data value.

Additionally, further work towards high-bandwidth vehicular communication networks can serve to ease the onboard memory constraints the SBB works under. Currently, the SBB is designed to record data over the course of a day or multiple days. However, high-speed vehicle-to-everything communication allowing for rapid data upload in real time would significantly reduce the data recording period of the SBB and possibly transform the SBB into a downstream application to be applied after data upload.

Finally, although our manuscript focuses on the application of the SBB to autonomous cars, the SBB pipeline can be adapted for intelligent data recording in other domains as well. Autonomous and semi-autonomous systems are being developed for truck, sea, and air transport to accommodate increased volume~\cite{pasha2020holistic,dulebenets2020adaptive,KAGANALBAYRAK2020101818} and to improve safety~\cite{trosterer2017transport}. As the onboard sensor suites of these mediums continue to increase in complexity and data bandwidth, intelligent event data recorders must be developed to store and manage valuable sensor data.
\section{Conclusions} \label{sec:conclusions}
This paper has proposed a novel Smart Black Box (SBB) data processing pipeline that uses video anomaly detection and online action detection to efficiently record large-scale high-value video data. We have addressed storage and value estimation problems the SBB will face with real world data, made adjustments in data classification and value estimation accordingly, and presented results on a large-scale real-world driving video dataset. Value estimation is changed from an entirely information measure-based method using pre-defined EOIs to use a combination of video anomaly detection and online action detection capable of detecting more generalized EOIs. Observed decision differences between normal and anomalous data indicate that SBB value estimation can distinguish normal and anomalous frames. In experiments, a 24.4\% increase in the anomalous-to-normal memory ratio was achieved compared to the raw data, in addition to a $\sim25\%$ to $\sim100\%$ increase in the anomalous-to-normal count ratio. However, we also noted that SBB performance increases significantly given ground-truth anomaly labels, suggesting that improved methods for general EOI detection will further improve the SBB utility. Future research in anomaly detection using sensor fusion, high-bandwidth vehicular communication networks, and intelligent event data recorders for other domains of transport can help realize prioritized data recording and storage for intelligent transportation systems.


\vspace{6pt} 

\authorcontributions{conceptualization, Y.Y. and E.A.; methodology, R.F. and Y.Y.; software, R.F. and Y.Y.; validation, R.F. and Y.Y.; formal analysis, R.F. and Y.Y.; investigation, R.F. and Y.Y.; resources, E.A.; data curation, R.F.; writing--original draft preparation, R.F.; writing--review and editing, Y.Y. and E.A.; visualization, R.F.; supervision, E.A.; project administration, Y.Y. and E.A.; funding acquisition, E.A.}

\funding{This research has been supported by  a grant from Ford Motor Company via the Ford-UM Alliance under Award N028603 and the National Science 
Foundation Award Number CNS 1544844.}

\conflictsofinterest{The authors declare no conflict of interest.} 


\reftitle{References}
\externalbibliography{yes}
\bibliography{ref}
\end{document}